\documentstyle[12pt,oldlfont]{article}
\normalbaselines
\begin{document}
\bibliographystyle{unsrt}

\begin{center}
{\Large {\bf PHOTON STATISTICS FOR MULTIMODE SQUEEZED SCHR\"ODINGER
CAT STATES}}
\end{center}

\bigskip

\begin{center}
{\bf V. I. Man'ko}
\end{center}

\smallskip
\begin{center}
{\em Lebedev Physical Institute, Leninsky pr., 53\\
Moscow 117924, Russia\\}

\end{center}

\bigskip

\begin{abstract}
{\small Particle distributions in squeezed states, even and odd
coherent states are given in terms of multivariable Hermite
polynomials. The Q--function and Wigner function for nonclassical
field states are discussed.}
\end{abstract}

\bigskip

\section{Introduction}

The aim of the talk is to give a review of such nonclassical states of
light as squeezed states \cite{hol}, \cite{yuen}, correlated states
\cite{kur}, even and odd coherent states \cite{dod74} (Schr\"odinger
cat states \cite{yur-sto}), squeezed Schr\"odinger cat states
\cite{nikon}. First we discuss the photon distribution function for
the generalized correlated states \cite{sudar} of multimode light.
For finding solutions of Schr\"odinger equation with time--dependent
Hamiltonians which are generic quadratic forms in position and momentum
operators the integrals of motion which are linear forms have been
constructed in Refs. \cite{mal70}, \cite{Malkinmankotrifonov69} and
\cite{tri70}. Such integrals of motion have been analyzed and applied to
general problems of quantum mechanics and statistics in Refs.
\cite{mal79},\cite{dod89}.

The nonstationary Hamiltonians are appropriate models for the physical
conditions in which nonclassical states of fields (photons, phonons,
pions, {\it etc.}) may be created. Following \cite{olga} we will
give the result for photon distribution function of generic mixed
squeezed and correlated Gaussian state. Initially the state is taken to
be standard coherent one (partial case of such state is photon vacuum)
and, for example, due to nonstationary Casimir effect it becomes
multimode mixed correlated state.

\section{Multimode Hermite Polynomials and Mixed Correlated Light}

The most general mixed squeezed state of the N--mode light with a
Gaussian density operator $~\hat{\varrho }$ is described by
the Wigner function $~W({\bf p},{\bf q})$ of the generic Gaussian form,
\begin{equation}
W({\bf p},{\bf q})=(\det {\bf M})^{-\frac 12}\exp\left
[-\frac 12({\bf Q}-<{\bf Q}>){\bf M}^{-1}({\bf Q}-<{\bf Q}>)\right],
\end{equation}
where 2N--dimensional vector $~{\bf Q}=({\bf p},{\bf q})$ consists
of N components $~p_1,~\ldots ,~p_N$ and N components
$~q_1,~\ldots ,~q_N$,
operators $~\hat {{\bf p}}$ and $~\hat {{\bf q}}$ being the
quadrature components of the photon creation
$~\hat {{\bf a}}\dag$ and annihilation $~\hat {{\bf a}}$
operators (we use dimensionless variables and assume $~\hbar =1$):
\begin{eqnarray}
\hat {{\bf p}}&=&\frac {\hat {{\bf a}}
-\hat {{\bf a}}\dag}{i\sqrt {2}},\nonumber\\
\hat {{\bf q}}&=&\frac {\hat {{\bf a}}
+\hat {{\bf a}}\dag}{\sqrt {2}}.
\end{eqnarray}
2N parameters $~<p_i>$ and $~<q_i>$, $~i=1,2,\ldots ,N$, combined
into vector $~<{\bf Q}{\bf >}$, are the average values of the
quadratures,
\begin{eqnarray}
<{\bf p}>&=&\mbox{Tr}~\hat{\varrho}\hat {{\bf p}},\nonumber\\
<{\bf q}>&=&\mbox{Tr}~\hat{\varrho}\hat {{\bf q}}.
\end{eqnarray}
A real symmetric dispersion matrix $~{\bf M}$ consists of 2N$^2$+N
variances
\begin{equation}
{\cal M}_{\alpha\beta}=\frac 12\left\langle\hat Q_{\alpha}\hat
Q_{\beta}+\hat Q_{\beta}\hat Q_{\alpha}\right\rangle -\left\langle
\hat Q_{\alpha}\right\rangle\left\langle\hat Q_{\beta}\right\rangle
,~~~~~~~~~~~\alpha ,\beta =1,2,\ldots ,2N.
\end{equation}
They obey certain constraints, which are nothing but the
generalized uncertainty relations \cite{dod89}.

The photon distribution function of this state
\begin{equation}
{\cal P}_{{\bf n}}=\mbox{Tr}~\hat{\varrho }|{\bf n}><{\bf n}|,
{}~~~~~~~{\bf n}=(n_1,n_2,\ldots ,n_N),
\end{equation}
where the state $~|{\bf n}>$ is photon number state, which was
calculated in \cite{olga}, and it is
\begin{equation}
{\cal P}_{{\bf n}}={\cal P}_0\frac {H_{{\bf n}{\bf n}}^{
\{{\bf R}\}}({\bf y})}{{\bf n}!}.
\end{equation}
The function $~H_{{\bf n}{\bf n}}^{\{{\bf R}\}}({\bf y})$ is
multidimensional Hermite polynomial. The probability to have no
photons is
\begin{equation}
{\cal P}_0=\left[\det\left({\bf M}+\frac 12{\bf I}_{2N}\right)\right
]^{-\frac 12}\exp\left[-<{\bf Q}>\left(2{\bf M}+{\bf I}_{2N}\right
)^{-1}<{\bf Q}>\right],
\end{equation}
where we introduced the matrix
\begin{equation}
{\bf R}=2{\bf U}^{\dag }(1+2{\bf M})^{-1}{\bf U}^{*}-\sigma _{Nx},
\end{equation}
and the matrix
\begin{equation}
\sigma _{Nx}=\left(\begin{array}{cc}
0&{\bf I}_N\\
{\bf I}_N&0\end{array}\right).
\end{equation}
The argument of Hermite polynomial is
\begin{equation}
{\bf y}=2{\bf U}^t({\bf I}_{2N}-2{\bf M})^{-1}<{\bf Q}>,
\end{equation}
and the 2N--dimensional unitary matrix
\begin{equation}
{\bf U}=\frac 1{\sqrt {2}}\left(\begin{array}{cc}
-i{\bf I}_N&i{\bf I}_N\\
{\bf I}_N&{\bf I}_N\end{array}\right)
\end{equation}
is introduced, in which $~{\bf I}_N$ is the N$\times $N identity
matrix. Also we use the notation
$${\bf n}!=n_1!n_2!...n_N!.$$

The mean photon number for j--th mode is expressed in terms of
photon quadrature means and dispersions
\begin{eqnarray}
<n_j>=\frac 12(\sigma_{p_jp_j}+\sigma_{q_jq_j}-1)
+\frac 12(<p_j>^2+<q_j^2>).
\end{eqnarray}
We introduce a complex 2N--vector $~{\bf B}=(\beta _{1},~\beta _{2},
{}~\ldots ,~\beta _{N},~\beta _{1}^{*},~\beta _{2}^{*},~\ldots ,
{}~\beta _{N}^{*})$. Then the Q--function \cite{hus40} is
the diagonal matrix element of the density operator in the coherent
state basis $~|~\beta _{1},~\beta _{2},~\ldots ,~\beta _{N}>.$
This function is the generating function for the matrix elements of the
density operator in the Fock basis $~|${\bf n}$>$ which has been calculated
in \cite{semen}. In notations corresponding to the Wigner function (1)
the Q--function is
\begin{equation}
Q({\bf B})={\cal P}_{0}\exp [-\frac {1}{2}{\bf B}(R+\sigma _{Nx}){\bf B}+
{\bf B}R{\bf y}].
\end{equation}
Thus, if the Wigner function (1) is given one has the Q--function. Also,
if one has the Q--function (13), i. e. the matrix $~R$ and the vector {\bf y},
the Wigner function may be obtained due to the relations
\begin{eqnarray}
{\bf M}&=&{\bf U}^{*}(R+\sigma _{Nx})^{-1}{\bf U}^{\dag }
-\frac {1}{2},\nonumber\\
<{\bf Q}>&=&{\bf U}^{*}[1-(R+\sigma _{Nx})^{-1}\sigma _{Nx}]{\bf y}.
\end{eqnarray}
For pure squeezed and correlated state with the wave function
\begin{equation}
\Psi =N\exp [-{\bf x}m{\bf x}+{\bf c}{\bf x}],
\end{equation}
where
\begin{equation}
N=[\det (m+m^{*})]^{1/4}\pi ^{-N/4}\exp \{\frac {1}{4}({\bf c}
+{\bf c}^{*})(m+m^{*})^{-1}({\bf c}+{\bf c}^{*})\},
\end{equation}
the symmetric 2N$\times $2N--matrix $~R$ determining Q--function
has the block--diagonal form
\begin{equation}
R=\left( \begin{array}{clcr}r&0\\
0&r^{*}\end{array}\right).
\end{equation}
The N$\times $N--matrix $~r$ is expressed in terms of symmetric matrix
$~m$
\begin{equation}
r^{*}=1-(m+1/2)^{-1},
\end{equation}
and the 2N--vector $~{\bf y}=({\bf Y,Y}^{*})$ is given by the
relation
\begin{equation}
{\bf Y}^{*}=\frac {1}{\sqrt 2}(m-1/2)^{-1}{\bf c}.
\end{equation}
The corresponding blocks of the dispersion matrix
\begin{equation}
{\bf M}=\left( \begin{array}{clcr}\sigma _{\bf pp}&\sigma _{\bf pq}\\
\widetilde \sigma _{\bf pq}&\sigma _{\bf qq}\end{array}\right)
\end{equation}
are
\begin{eqnarray}
\sigma _{\bf pp}&=&2(m^{-1}+m^{*-1})^{-1},\nonumber\\
\sigma _{\bf qq}&=&\frac {1}{2}(m+m^{*})^{-1},\nonumber\\
\sigma _{\bf pq}&=&\frac {i}{2}(m-m^{*})(m+m^{*})^{-1}.
\end{eqnarray}
The probability to have no photons is
$$
P_{0}=\frac {[\det (m+m^{*})]^{1/2}}{|\det (m+1/2)|}$$
\begin{equation}
\otimes \exp \{\frac {1}{2}({\bf c}+{\bf c}^{*})(m+m^{*})^{-1}({\bf c}
+{\bf c}^{*})+\frac {1}{4}[{\bf c}(m+1/2)^{-1}{\bf c}+{\bf c}^{*}(m^{*}+
1/2)^{-1}{\bf c}^{*}]\}.
\end{equation}
The multivariable Hermite polynomials describe the photon distribution
function for the multimode mixed and pure correlated light \cite{olga},
\cite{md94}, \cite{dodon94}.The nonclassical state of the light may be
created due to nonstationary Casimir effect \cite{jslr}, and the
Husimi oscillator is the model to describe the behaviour of
the squeezed and correlated photons.

\section{Multimode Even and Odd Coherent States}

We define the multimode even and odd coherent states (Schr\"odinger cat
male states and Schr\"odinger cat female states, respectively)
as \cite{ans}
\begin{equation}
\mid {\bf A_{\pm}}>=N_{\pm} (\mid {\bf A}> \pm \mid -{\bf A}>),
\end{equation}
where the multimode coherent state $\mid {\bf A}>$ is
\begin{equation}
\mid {\bf A}>=\mid \alpha_{1},~\alpha_{2},~\ldots ,~\alpha_{n}>
=D({\bf A}) \mid {\bf 0}>,\nonumber\\
\end{equation}
and the multimode coherent state is created from multimode vacuum state
$\mid {\bf 0}>$ by the multimode displacement operator $D({\bf A})$.
The definition of multimode even and odd coherent states is the obvious
generalization of the one--mode even and odd coherent state given in
\cite{dod74}, \cite{mal79}.
The normalization constants for multimode even and odd coherent states
become
\begin{eqnarray}
N_{+}&=&\frac{e^{\frac{\mid {\bf A} \mid^{2}}{2}}}
{2 \sqrt{\cosh\mid {\bf A}\mid^{2}}},\nonumber\\
N_{-}&=&\frac{e^{\frac{\mid {\bf A} \mid^{2}}{2}}}
{2 \sqrt{\sinh\mid {\bf A}\mid^{2}}},
\end{eqnarray}
where $~{\bf A}=(~\alpha_{1},~\alpha_{2},~\ldots,~\alpha_{n})$ is a
complex vector and its modulus is
\begin{equation}
\mid {\bf A} \mid^{2}=\mid \alpha_{1} \mid^{2}+\mid \alpha_{2} \mid^{2}
+\ldots +\mid \alpha_{n} \mid^{2}=\sum_{m=1}^{n}\mid \alpha_{m} \mid^{2}.
\nonumber\\
\end{equation}
Such multimode even and odd coherent states can be decomposed into multimode
number states as
\begin{equation}
\mid {\bf A_{\pm}}>=N_{\pm}\sum_{{\bf n}}
\frac{e^{-\frac {1}{2}\mid {\bf A}\mid^2}
\alpha_{1}^{n_{1}}\ldots \alpha_{n}^{n_{n}}}{\sqrt{n_{1}!}
\ldots \sqrt{n_{n}!}}
(1\pm(-1)^{n_{1}+n_{2}+\ldots n_{n}}) \mid {\bf n}>,\nonumber\\
\end{equation}
where $\mid {\bf n}>=~\mid n_{1},~n_{2},~\ldots ,~n_{n}>$ is the
multimode number state. Also from Eq. (23), we can derive an important
relation for the multimode even and odd coherent states, namely,
\begin{eqnarray}
a_{i} \mid {\bf A_{+}}>&=&\alpha_{i} \sqrt{\tanh\mid {\bf A}\mid^{2}}
\mid {\bf A_{-}}>,\nonumber\\
a_{i} \mid {\bf A_{-}}>&=&\alpha_{i} \sqrt{\coth\mid {\bf A}\mid^{2}}
\mid {\bf A_{+}}>.\nonumber\\
\end{eqnarray}
The probability of finding {\bf n} photons in multimode even and odd
coherent states can be worked out with the help of Eq. (27)
\begin{eqnarray}
P_{+} ({\bf n})&=&\frac{\mid \alpha_{1} \mid^{2n_{1}}
\mid \alpha_{2} \mid^{2n_{2}}\ldots \mid \alpha_{n} \mid^{2n_{n}} }
{(n_{1}!)(n_{2}!)\ldots (n_{n}!)\cosh\mid {\bf A}\mid^{2}},
{}~~~n_{1}+n_{2}+\ldots +n_{n}=2k,\nonumber\\
 P_{-} ({\bf n})&=&\frac{\mid \alpha_{1} \mid^{2n_{1}}
\mid \alpha_{2} \mid^{2n_{2}}\ldots \mid \alpha_{n} \mid^{2n_{n}} }
{(n_{1}!)(n_{2}!)\ldots (n_{n}!)\sinh\mid {\bf A}\mid^{2}},
{}~~~n_{1}+n_{2}+\ldots +n_{n}=2k+1.\nonumber\\
\end{eqnarray}
Multimode coherent states are the product of independent coherent states
of each mode, and photon distribution function is the product of independent
Poissonian distribution functions. But in the present case of multimode
even and odd coherent states we cannot factorize their multimode
photon distribution functions due to the presence of the nonfactorizable
$\cosh\mid {\bf A}\mid^{2}$ and $\sinh\mid {\bf A}\mid^{2}$.
This fact implies the phenomenon of statistical dependences of different modes
of these states on each other.

In order to describe the properties of the distribution functions from
Eq. (29) we will calculate the symmetric 2N$\times$2N dispersion matrix
for multimode field quadrature components. For even and odd coherent states
we have
\begin{equation}
<{\bf A}_{\pm} \mid a_{i}a_{k} \mid {\bf A}_{\pm}>=\alpha_{i}\alpha_{k},
\end{equation}
and complex conjugate values of the above equation for
$<{\bf A}_{\pm} \mid a_{i}^{\dag}a_{k}^{\dag} \mid {\bf A}_{\pm}>$.
Since the quantity $<{\bf A}_{\pm} \mid a_{i} \mid {\bf A}_{\pm}>$ is equal to
zero the above equation gives two N$\times$N blocks of the dispersion matrix.
For other two N$\times$N blocks of this matrix
we have for the multimode even coherent states
\begin{equation}
\sigma_{(a_{i}^{\dag}a_{k})}^{+}=<{\bf A}_{+} \mid
\frac{1}{2}(a_{i}^{\dag}a_{k}
+a_{k}a_{i}^{\dag})\mid {\bf A_{+}}>=\alpha_{i}^{*}\alpha_{k}
\tanh\mid {\bf A} \mid^{2}+\frac{1}{2} \delta_{ik},\nonumber\\
\end{equation}
and for multimode odd coherent states
\begin{equation}
\sigma_{(a_{i}^{\dag}a_{k})}^{-}=<{\bf A}_{-} \mid
\frac{1}{2}(a_{i}^{\dag}a_{k}
+a_{k}a_{i}^{\dag})\mid {\bf A_{-}}>=\alpha_{i}^{*}\alpha_{k}
\coth\mid {\bf A} \mid^{2}+\frac{1}{2} \delta_{ik}.\nonumber\\
\end{equation}
For the dispersion matrix, the mean values of the photon
numbers $n_{i}=a_{i}^{\dag}a_{i}$ for multimode even and odd coherent states
are the following
\begin{eqnarray}
<{\bf A_{+}}\mid n_{i} \mid {\bf A_{+}}>&=&\mid \alpha_{i}\mid^{2}
\tanh\mid {\bf A}\mid^{2},\nonumber\\
<{\bf A_{-}}\mid n_{i} \mid {\bf A_{-}}>&=&\mid \alpha_{i}\mid^{2}
\coth\mid {\bf A}\mid^{2}.
\end{eqnarray}
Taking into account the above equation the symmetric N$\times$N dispersion
matrices for photon number operators can be obtained from the above
given distribution functions for multimode even and odd coherent states.
By defining
\begin{equation}
\sigma_{ik}^{\pm}=<{\bf A}_{\pm}\mid n_{i}n_{k}\mid,
{\bf A}_{\pm}>,
\end{equation}
the corresponding  expressions in such states are
\begin{eqnarray}
\sigma _{ik}^{+}&=&\mid \alpha _{i} \mid ^{2}\mid \alpha _{k}\mid ^{2}
\mbox {sech }^{2}\mid {\bf A}\mid ^{2}+\mid \alpha _{i}\mid ^{2}
\tanh \mid {\bf A}\mid ^{2}\delta _{ik},\nonumber\\
\sigma _{ik}^{-}&=&-\mid \alpha _{i}\mid ^{2}\mid \alpha _{k}\mid^{2}
\mbox {csch }^{2}\mid {\bf A}\mid ^{2}+\mid \alpha _{i}\mid ^{2}
\coth \mid {\bf A}\mid ^{2}\delta _{ik}.
\end{eqnarray}
As the nondiagonal matrix elements of the dispersion density matrix
are not equal to zero so we can predict
that  different modes of these states are
correlated with each other. In other words, as we have mentioned before,
there exist some statistical dependences of different modes on each other.

Another interesting property for the multimode even and odd coherent states
is the Q--function, and it can be obtained in the following manner. First
of all the density matrices for the multimode even and odd coherent states
are
\begin{equation}
\rho_{\pm}=\mid {\bf A_{\pm}}><{\bf A_{\pm}} \mid,
\end{equation}
then the Q--function can be calculated as
\begin{eqnarray}
 Q_{+}({\bf B},{\bf B}^{*})&=&<{\bf B} \mid \rho_{+}
\mid {\bf B}>\nonumber\\
&=&4N_{+}^{2}e^{-(\mid {\bf A}\mid ^{2}
+\mid {\bf B}\mid ^{2})}\mid
\cosh({\bf A {\bf B}^{*}})\mid ^{2}\nonumber\\
Q_{-}({\bf B},{\bf B}^{*})&=&<{\bf B} \mid \rho_{-}
\mid {\bf B}>\nonumber\\
&=&4N_{-}^{2}e^{-(\mid {\bf A}\mid ^{2}
+\mid {\bf B}\mid ^{2})}\mid \sinh({\bf A
{\bf B}^{*}})\mid ^{2},
\end{eqnarray}
where $\mid {\bf B}>=\mid \beta_{1},~\beta_{2},~\ldots ,~\beta_{n}>$
is another multimode coherent state with multimode eigenvalue
${\bf B}=(\beta_{1},~\beta_{2},~\ldots ,~\beta_{n})$.
We call these functions for even and odd coherent states
as the Q--functions for the Schr\"odinger cat states.
The Q--function for single--mode odd coherent state shows the
crater type behaviour for small values of the quantity
$\mid \alpha \mid$ and for its larger values the Q--function begins to
split into two peaks in a similar manner as in case of even coherent
states \cite{ans}.

The Wigner function for the multimode coherent states is \cite{dod89}
\begin{equation}
W_{{\bf A,B}}=2^{N}\exp[-2{\bf Z Z^{*}}+2{\bf A Z^{*}}
+2{\bf B^{*}Z}-{\bf AB^{*}}-\frac{\mid {\bf A}\mid^{2}}{2}
-\frac{\mid {\bf B}\mid^{2}}{2}],
\nonumber\\
\end{equation}
where
\begin{equation}
{\bf Z}=\frac{{\bf q+i p}}{\sqrt{2}}.
\end{equation}
For even and odd coherent states the Wigner function is
\begin{eqnarray}
W_{{\bf A}_{\pm}}({\bf q,p})
&=&\mid N_{\pm}\mid^{2}[W_{{\bf (A,B=A)}}({\bf q,p})\pm
W_{{\bf (A,B=-A)}}({\bf q,p})\nonumber\\
&\pm &W_{{\bf (-A,B=A)}}({\bf q,p})+W_{({\bf -A,B=-A)}}({\bf q,p})],
\end{eqnarray}
where the explicit forms of $N_{\pm}$ are given in Eq. (25).
For multimode case we use the following notations
\begin{eqnarray}
{\bf AZ^{*}}&=&\alpha_{1}Z_{1}^{*}+\alpha_{2}Z_{2}^{*}
+\ldots \alpha_{n}Z_{n}^{*},
\nonumber\\
{\bf ZZ^{*}}&=&Z_{1}Z_{1}^{*}+Z_{2}Z_{2}^{*}+\ldots +Z_{n}Z_{n}^{*}.
\end{eqnarray}

The photon distribution function gives the probability of finding
2k photons for two--mode even coherent state, and is defined as
\begin{equation}
P_{+}(2k)=\frac{(\mid \alpha_{1} \mid^{2}+\mid \alpha_{2} \mid^{2})^{2k}}
{(2k)!\cosh(\mid \alpha_{1} \mid^{2}+\mid \alpha_{2} \mid^{2})},
\end{equation}
where $2k=n_{1}+n_{2}$, for both $n_{1}$ and $n_{2}$ to be even or odd
numbers. Similarly for two--mode odd coherent state it gives the
probability of finding 2k+1 photons
\begin{equation}
P_{-}(2k+1)=\frac{(\mid \alpha_{1} \mid^{2}
+\mid \alpha_{2} \mid^{2})^{2k+1}}
{(2k+1)!\sinh(\mid \alpha_{1} \mid^{2}+\mid \alpha_{2} \mid^{2})}.
\end{equation}
For this case, $~n_{1}$ is even and $~n_{2}$ is odd number, or vice versa.
For single--mode case the photon distribution
functions demonstrate super and sub--Poissonian properties for even and odd
coherent states, respectively. The same conclusions may be drawn for
two--mode (and multimode) even and odd coherent states.

\section{Some Relations for Wigner and Q--functions}

The Wigner function of a system  $~W({\bf p,q})=~W({\bf Q})$ is expressed
in terms of density matrix in coordinate representation as (see, for example,
\cite{dod89})
\begin{equation}
W({\bf p,q})=\int \rho ({\bf q}+\frac {\bf u}{2},~{\bf q}-\frac {\bf u}{2})
\exp (-i{\bf p}{\bf u})~d{\bf u}.
\end{equation}
The inverse transform is
\begin{equation}
\rho ({\bf x,x'})=\frac {1}{(2\pi )^{N}}\int W(\frac {{\bf x}+{\bf x'}}{2},
{}~{\bf p})\exp [i{\bf p}({\bf x}-{\bf x'})]~d{\bf p}.
\end{equation}
The Q--function is expressed in terms of the Wigner function through the
3N--dimensional integral transform
\begin{equation}
Q({\bf B})=\frac {1}{(2\pi )^{N}}\int \Phi _{\bf B}({\bf x,x',p})W(\frac
{{\bf x}+{\bf x'}}{2},{\bf p})~d{\bf x}~d{\bf x'}~d{\bf p}
\end{equation}
with the kernel
\begin{equation}
\Phi _{\bf B}({\bf x,x',p})=\pi ^{-N/2}\exp [i{\bf p}({\bf x}-{\bf x'})
-\frac {1}{2}{\bf B}(\sigma _{Nx}+{\bf I}_{2N}){\bf B}
-\frac {{\bf X}^{2}}{2}+{\sqrt 2}{\bf B}\sigma _{Nx}{\bf X}],
\end{equation}
where the 2N--vector $~{\bf X}=({\bf x,x'})$ is introduced. If one has the
Q--function the Wigner function is given by the integral transform
\begin{equation}
W({\bf p,q})=\frac {1}{\pi ^{2N}}\int \{\prod _{k=1}^{N}d^{2}\beta _{k}
{}~d^{2}\gamma _{k}~du_{k}\widetilde \Phi _{k}(u_{k},\widetilde {\bf B})\}
Q(\widetilde {\bf B}),
\end{equation}
where the argument of the Q--function $~{\bf B}$ is replaced by the
2N--vector with complex components
$$\widetilde {\bf B}=(~\beta _{1},~\beta _{2},~\ldots ,~\beta _{N},
{}~\gamma _{1}^{*},~\gamma _{2}^{*},~\ldots ,~\gamma _{N}^{*}),$$
and the kernel has the form
\begin{eqnarray}
\widetilde \Phi _{k}(u_{k},\widetilde {\bf B})&=&\pi ^{-1/2}
\exp [-|\beta _{\dot k}|^{2}
-|\gamma _{\dot k}|^{2}+{\sqrt 2}(q_{\dot k}
+\frac {u_{\dot k}}{2})\beta _{\dot k}+{\sqrt 2}(q_{\dot k}
-\frac {u_{\dot k}}{2})\gamma _{\dot k}^{*}\nonumber\\
&-&\frac {1}{2}(q_{k}+\frac {u_{k}}{2})^{2}-\frac {1}{2}(q_{k}
-\frac {u_{k}}{2})^{2}-\frac {\beta _{\dot k}^{2}}{2}
-\frac {\gamma _{\dot k}^{*2}}{2}
-ip_{\dot k}u_{\dot k}+\gamma _{\dot k}\beta _{\dot k}^{*}].
\end{eqnarray}
The density matrix in coordinate representation is related to the
Q--function
\begin{equation}
\rho ({\bf x,x'})=\pi ^{-2N}\int \{\prod _{k=1}^{N}d^{2}\beta _{k}
{}~d^{2}\gamma _{k}\phi_{k} (\widetilde {\bf B})
\exp [-\frac {1}{2}(x_{k}^{2}+x_{k}^{'2})]\}
Q(\widetilde {\bf B}),
\end{equation}
where the kernel of the transform is
\begin{equation}
\phi _{k}(\widetilde {\bf B})=\pi ^{-1/2}
\exp [-|\beta _{k}|^{2}
-|\gamma _{k}|^{2}+{\sqrt 2}x_{k}\beta _{k}
+{\sqrt 2}x'_{k}\gamma _{k}^{*}
-\frac {\beta _{k}^{2}}{2}-\frac {\gamma _{k}^{*2}}{2}
+\gamma _{k}\beta _{k}^{*}].
\end{equation}
The evolution of the Wigner function and Q--function for systems with
quadratic Hamiltonians for any state is given by the following
prescription. Given the Wigner function $~W({\bf p,q},t=0)$
for the initial time $~t=0.$ Then the Wigner function for the time $~t$
is obtained by the replacement
$$W({\bf p,q},t)=~W({\bf p}(t),~{\bf q}(t),~t=0),$$
where the time--dependent arguments are the linear integrals of motion
of the quadratic system found in \cite{tri73}, \cite{dod89}. The same
ansatz is used for the Q--function. Namely, given the Q--function
of the quadratic system $~Q({\bf B},~(t=0))$ for the initial
time $~t=0.$ Then the Q--function for the time $~t$ is given by
the replacement
$$Q({\bf B},~t)=~Q({\bf B}(t),~t=0),$$
where the 2N--vector $~{\bf B}(t)$ is the integral of motion linear in
annihilation and creation operators found in \cite{tri73}, \cite{dod89}.
This ansatz follows from the statement that the density operator of the
Hamiltonian system is the integral of motion, and its matrix elements in
any basis must depend on the appropriate integrals of motion. In particular,
the Wigner function and Q--function depend just on the linear invariants
found in \cite{tri73}, \cite{dod89}.

\section{Multivariable Hermite Polynomials}
For parametric forced oscillator the transition amplitude between
its energy levels has been calculated as overlap integral of two generic
Hermite polynomials with a Gaussian function (Frank--Condon factor) and
expressed in terms of Hermite polynomials of two variables \cite{mal70}.
For N--mode parametric oscillator the analogous amplitude has been
expressed in terms of Hermite polynomials of 2N variables, i. e. the
overlap integral of two generic Hermite polynomials of N variables with
a Gaussian function (Frank--Condon factor for a polyatomic molecule) has
been evaluated in \cite{tri73}. The corresponding result uses the formula\\
$$({\bf n}=n_{1},n_{2},\ldots n_{N},~~~~~{\bf m}=m_{1},m_{2},\ldots
m_{N},~~~~~m_{i},n_{i}=0,1,\ldots),$$
\begin{equation}
\int H_{{\bf n}}^{\{R\}}({\bf x})H_{{\bf m}}^{\{r\}}(\Lambda {\bf x}
+{\bf d})\exp (-{\bf x}m{\bf x}+{\bf c}{\bf x})~d{\bf x}
=\frac {\pi ^{N/2}}{\sqrt {\det m}}\exp (\frac {1}{4}{\bf c}
m^{-1}{\bf c})H_{{\bf mn}}^{\{\rho \}}({\bf y}),
\end{equation}
where the symmetric 2N$\times $2N--matrix
\begin{equation}
\rho=\left( \begin{array}{clcr}R_{1}&R_{12}\\
\widetilde R_{12}&R_{2}\end{array}\right)
\end{equation}
with N$\times $N--blocks $~R_{1},~R_{2},~R_{12}$ is expressed in terms
of symmetric N$\times $N--matrices $~R,~r,~m$ and N$\times $N--matrix
$~\Lambda $ in the form
\begin{eqnarray}
R_{1}&=&R-\frac {1}{2}Rm^{-1}R,\nonumber\\
R_{2}&=&r-\frac {1}{2}r\Lambda m^{-1}\tilde \Lambda r,\nonumber\\
\widetilde R_{12}&=&-\frac {1}{2}r\Lambda m^{-1}R.
\end{eqnarray}
Here the matrix $~\widetilde \Lambda $ is transposed matrix $~\Lambda $
and $~\widetilde R_{12}$ is transposed matrix $~R_{12}.$ The 2N--vector
$~{\bf y}$ is expressed in terms of N--vectors $~{\bf c}$ and $~{\bf d}$
in the form
\begin{equation}
{\bf y}=\rho ^{-1}\left( \begin{array}{c}{\bf y}_{1}\\
{\bf y}_{2}\end{array}\right ),
\end{equation}
where the N--vectors $~{\bf y}_{1}$ and $~{\bf y}_{2}$ are
\begin{eqnarray}
{\bf y}_{1}&=&\frac {1}{4}(Rm^{-1}+m^{-1}R){\bf c}\nonumber\\
{\bf y}_{2}&=&\frac {1}{4}(r\Lambda m^{-1}
+m^{-1}\tilde \Lambda r){\bf c}+r{\bf d}.
\end{eqnarray}
For matrices $~R=2,~r=2$ the above formula (52) yields
$$\int \{\prod _{i=1}^{N}H_{n_{i}}(x_{i})H_{m_{i}}
(\sum _{k=1}^{N}\Lambda _{ik}x_{k}+d_{i})\}\exp (-{\bf x}m{\bf x}
+{\bf c}{\bf x})~d{\bf x}$$
\begin{equation}
=\frac {\pi ^{N/2}}{\sqrt {\det m}}
\exp (\frac {1}{4}{\bf c}m^{-1}{\bf c})H_{{\bf mn}}^{\{\rho \}}({\bf y}),
\end{equation}
with N$\times $N--blocks $~R_{1},~R_{2},~R_{12}$ expressed in terms of
N$\times $N--matrices $~m$ and $~\Lambda $
in the form
\begin{eqnarray}
R_{1}&=&2(1-m^{-1}),\nonumber\\
R_{2}&=&2(1-\Lambda m^{-1}\tilde \Lambda ),\nonumber\\
\widetilde R_{12}&=&-2\Lambda m^{-1}.
\end{eqnarray}
The 2N--vector
$~{\bf y}$ is expressed in terms of N--vectors $~{\bf c}$ and $~{\bf d}$
in the form (55) with
\begin{eqnarray}
{\bf y}_{1}&=&m^{-1}{\bf c},\nonumber\\
{\bf y}_{2}&=&\frac {1}{2}(\Lambda m^{-1}
+m^{-1}\tilde \Lambda ){\bf c}+2{\bf d}.
\end{eqnarray}
If the symmetric matrix $~\rho $ has the block--diagonal structure
$$
\rho=\left( \begin{array}{clcr}R_{1}&0\\
0&R_{2}\end{array}\right)
$$
with the symmetric S$\times $S--matrix $~R_{1}$ and the symmetric
(2N-S)$\times $(2N-S)--matrix $~R_{2}$ the multivariable Hermite
polynomial is represented as the product of two Hermite polynomials
depending on S and 2N-S variables, respectively,
$$
H_{\bf k}^{\{{\bf \rho }\}}({\bf y})
=H_{{\bf n}_{S}}^{\{{\bf R}_{1}\}}({\bf y}_{1})
H_{{\bf n}_{2N-S}}^{\{{\bf R}_{2}\}}({\bf y}_{2}),
$$
where the 2N--vector $~{\bf y}$ has the vector--components
$${\bf y}=(~{\bf y}_{1},~{\bf y}_{2}),$$
and 2N--vector $~{\bf k}$ has the components
$${\bf k}=({\bf n}_{S},~{\bf n}_{2N-S})=(~n_{1},~\ldots ,~n_{S},~n_{S
+1},~\ldots ,~n_{2N}).$$
The partial case of this relation is the relation for the Hermite
polynomials with the matrix $~R$ with complex conjugate blocks
$~R_{1}=~r,~~R_{2}=~r^{*}$, and complex conjugate vector--components
$~{\bf y}_{1}=~{\bf y}_{2}^{*}$
$$
H_{\bf k}^{\{{\bf \rho }\}}({\bf y})
=|H_{{\bf n}_{S}}^{\{{\bf R}_{1}\}}({\bf y}_{1})|^{2},~~~~~S=N.$$
The calculated integrals are important to evaluate the Green function
or density matrix for the systems with quadratic Hamiltonians. The
partial cases of multivariable Hermite polynomials determine some other
special functions \cite{md94}, \cite{dodon94}.

\section{Squeezing in Parametric Oscillator}
For the parametric oscillator with the Hamiltonian
\begin{equation}
H=-\frac {\partial ^{2}}{2\partial x^{2}}
+\frac {\omega ^{2}(t)x^{2}}{2},
\end{equation}
where we take $~\hbar=~m=~\omega (0)=~1$, there exists the
time--dependent integral of motion
\begin{equation}
A=\frac {i}{\sqrt 2}[\varepsilon (t)p-\dot \varepsilon (t)x],
\end{equation}
where
\begin{equation}
\ddot \varepsilon (t)+\omega ^{2}(t)\varepsilon (t)=0,~~~~~~
\varepsilon (0)=1,~~~~~~\dot \varepsilon (0)=i,
\end{equation}
satisfying the commutation relation
\begin{equation}
[A,~A\dag ]=1.
\end{equation}
It is easy to show that the packet solutions of the Schr\"odinger equation
may be introduced and interpreted as coherent states \cite{mal70}, since
they are eigenstates of the operator $~A$ (61), of the form
\begin{equation}
\Psi _{\alpha }(x,t)=\Psi _{0}(x,t)\exp \{-\frac {|\alpha |^{2}}{2}-
\frac {\alpha ^{2}\varepsilon ^{*}(t)}{2\varepsilon (t)}
+\frac {{\sqrt 2}\alpha x}{\varepsilon}\},
\end{equation}
where
\begin{equation}
\Psi _{0}(x,t)=\pi ^{-1/4}\varepsilon (t)^{-1/2}
\exp \frac {i\dot \varepsilon (t)x^{2}}{2\varepsilon (t)}
\end{equation}
is analog of the ground state of the oscillator and $~\alpha $ is a
complex number. The variances of the position and momentum of the
parametric oscillator in the state (65) are
\begin{equation}
\sigma _{x}=\frac {|\varepsilon (t)|^{2}}{2},~~~~~~\sigma _{p}
=\frac {|\dot \varepsilon (t)|^{2}}{2},
\end{equation}
and the correlation coefficient of the position and momentum has
the value corresponding to minimization of the Schr\"odinger uncertainty
relation \cite{schrod}
\begin{equation}
\sigma _{x}\sigma_{p}=\frac {1}{4}\frac {1}{1-r^{2}}.
\end{equation}
Another normalized solution to the Schr\"odinger equation
\begin{equation}
\Psi _{\alpha m}(x,t)=2N_{m}\Psi _{0}(x,t)\exp \{-\frac {|\alpha |^{2}}
{2}-\frac {\varepsilon ^{*}(t)\alpha ^{2}}{2\varepsilon (t)}\}\cosh
\frac {{\sqrt 2}\alpha x}{\varepsilon (t)},
\end{equation}
where
\begin{equation}
N_{m}=\frac {\exp (|\alpha |^{2}/2)}{2\sqrt {\cosh |\alpha |^{2}}}
\end{equation}
is the even coherent state \cite{dod74} (the Schr\"odinger cat male state).
The odd coherent state of the parametric oscillator (Schr\"odinger cat
female state)
\begin{equation}
\Psi _{\alpha f}(x,t)=2N_{f}\Psi _{0}(x,t)\exp \{-\frac {|\alpha |^{2}}
{2}-\frac {\varepsilon ^{*}(t)\alpha ^{2}}{2\varepsilon (t)}\}\sinh
\frac {\sqrt {2}\alpha x}{\varepsilon (t)},
\end{equation}
where
\begin{equation}
N_{f}=\frac {\exp (|\alpha |^{2}/2)}{2\sqrt {\sinh |\alpha |^{2}}},
\end{equation}
satisfies the Schr\"odinger equation and is the eigenstate of the
integral of motion $~A^{2}$ (as well as the even coherent state) with
the eigenvalue $~\alpha ^{2}$. These states are one--mode examples
of squeezed and correlated Schr\"odinger cat states constructed in
\cite{nikon}.

\section{Conclusion}
The discussed nonclassical states of the fields (photons, phonons, pions)
may be created in nonlinear interactions. The particle statistics with
squeezing and correlations of quadrature components may give an experimental
evidence of producing the new types of the field states. The
Schr\"odinger--like equations are used also in other branches of physics like
fiber optics. In \cite{renato} the Schr\"odinger--like equations has been
introduced to describe the charged particle beam in accelarator. The
approach described in the talk may be also applied in the classical physics
using the quantum--mechanical methods.


\begin{thebibliography}{99}

\bibitem{hol} J.N. Hollenhorst, {\em Phys. Rev.} {\bf D19} (1979) 1669.

\bibitem{yuen} H.P. Yuen, {\em Phys. Rev.} {\bf A13} (1976) 2226.

\bibitem{kur} V.V. Dodonov, E.V. Kurmyshev, and V.I. Man'ko,
{\em Phys. Lett.} {\bf A79} (1980) 150.

\bibitem{dod74} V.V. Dodonov, I.A. Malkin, and V.I. Man'ko, {\em Physica}
{\bf 72} (1974) 597.

\bibitem{yur-sto} B. Yurke and D. Stoler, {\em Phys. Rev. Lett.} {\bf 57}
(1986) 13.

\bibitem{nikon} V. V. Dodonov, V. I. Man'ko, and D. E. Nikonov, {\em
Even and odd coherent states (Schr\"odinger cat states) for multimode
parametric systems, Preprint} INFN-IY-NA-93/49, DSF-T-93/49 (1993);
{\em Phys. Rev.} {\bf A} (in press).

\bibitem{sudar} E. C. G. Sudarshan, {\em
Uncertainty relations, zero point energy and the linear
canonical group, Proc. of Second International Workshop on Squeezed
States and Uncertainty Relations, Moscow, 25-29 May, 1992}, eds. D.
Han, Y. S. Kim, and V.I.Man'ko (NASA Conference Publication {\bf 3219},
1993) p. 241.

\bibitem{mal70} I. A. Malkin and V. I. Man'ko, {\em Phys. Lett.}
{\bf A32} (1970) 243.

\bibitem{Malkinmankotrifonov69} I. A. Malkin, V. I. Man'ko, and D. A.
Trifonov, {\em Phys. Lett.} {\bf A30} (1969) 414.

\bibitem{tri70} I. A. Malkin, V. I. Man'ko, and D. A. Trifonov,
{\em Phys. Rev.} {\bf D2} (1970) 1371.

\bibitem{mal79} I. A. Malkin and V. I. Man'ko, {\em Dynamical Symmetries
and Coherent States of Quantum Systems} (Nauka Publishers,
Moscow, 1979) [in Russian].

\bibitem{dod89} V. V. Dodonov and V. I. Man'ko, {\em Invariants and
Evolution of Nonstationary Quantum Systems, Proc. of Lebedev Physics
Institute} {\bf 183}, ed. M. A. Markov (Nova Science, Commack, N. Y., 1989).

\bibitem{olga}  V. V. Dodonov, O. V. Man'ko, and V. I. Man'ko,
{\em Phys. Rev.} {\bf A50} (1994) 813.

\bibitem{hus40} K. Husimi, {\em Proc. Phys. Math. Soc. Japan} {\bf 22}
(1940) 264.

\bibitem{semen} V. V. Dodonov, V. I. Man'ko, and V. V. Semjonov,
{\em Nuovo Cim.} {\bf B83} (1984) 145.

\bibitem{md94} V. V. Dodonov and V. I. Man'ko, {\em J. Math. Phys.}
{\bf 35} (1994) 4277.

\bibitem{dodon94} V. V. Dodonov, {\em J. Phys.} {\bf A27} (1994) 6191.

\bibitem{jslr} V. I. Man'ko, {\it J. of Soviet Laser Research} (Plenum
Press, N. Y.) {\bf 12} (1991) N5.

\bibitem{ans} N. A. Ansari and V. I. Man'ko, {\em Phys. Rev.}
{\bf A50} (1994) 1942.

\bibitem{tri73} I. A. Malkin, V. I. Man'ko, and D. A. Trifonov, {\em
J. Math. Phys.} {\bf 14} (1973) 576.

\bibitem{schrod} E. Schr\"odinger, {\em Ber. Kgl. Akad. Wiss. Berlin}
{\bf 24} (1930) 296.

\bibitem{renato} R. Fedele and G. Miele, {\em Nuovo Cim.} {\bf D13}
(1991) 1527.

\end{thebibliography}
\end{document}